# A High Order Sliding Mode Control with PID Sliding Surface: Simulation on a Torpedo

Ahmed Rhif

Department of Electronics Engineering, High Institute of Applied Sciences and Technologies, Sousse, Tunisia
(Institut Supérieur des Sciences Appliquées et de Technologie de Sousse)
E-mail: ahmed.rhif@gmail.com

**ABSTRACT**

*Position and speed control of the torpedo present a real problem for the actuators because of the high level of the system non linearity and because of the external disturbances. The non linear systems control is based on several different approaches, among it the sliding mode control. The sliding mode control has proved its effectiveness through the different studies. The advantage that makes such an important approach is its robustness versus the disturbances and the model uncertainties. However, this approach implies a disadvantage which is the chattering phenomenon caused by the discontinuous part of this control and which can have a harmful effect on the actuators. This paper deals with the basic concepts, mathematics, and design aspects of a control for nonlinear systems that make the chattering effect lower. As solution to this problem we will adopt as a starting point the high order sliding mode approaches then the PID sliding surface. Simulation results show that this control strategy can attain excellent control performance with no chattering problem.*

**KEYWORDS**

*Sliding mode control, PID controller, chattering phenomenon, nonlinear system.*

## 1. INTRODUCTION

Modern torpedoes are the most effective marine weapons but they have a range much lower than the anti-ship missiles. Torpedoes are propelled engine equipped with an explosive charge, and sometimes with an internal guidance system that controls the direction, speed and depth. The typical shape of a torpedo is a cigar of 6 m long with a diameter of 50 cm and weighs one ton. The torpedoes are the main weapons of a submarine, but are also used by ships and by aircraft. They are increasingly wire-guided (cable several thousand meters connects the submarine making it possible to re-program or re-direct the machine according to the evolution of the target). However, most modern torpedoes can be completely autonomous. They have active sonar which makes them able to direct themselves to the target they have been designated prior to launch. Other types of torpedoes for example self-possessed, and especially during the second half of World War II, an acoustic sensor (passive sonar) allowed them to move to the noise emitted by the engines of the target. However, sometimes this kind of torpedo locks on the engine noise of the submarine pitcher, so the standard procedure was to dive low speed after such a shot. Modern torpedoes are powered by steam or electricity. The former have speeds ranging from 25 to 45 knots, and their scope ranges from 4 to 27 km. They consist of four elements: the warhead, the air section, rear section and tail section. The warhead is filled with explosive (181 to 363 kg). The steam-air section is about one third of the torpedo and contains compressed air and fuel tanks and water for the propulsion system. The rear section contains the turbine propulsion systems with the guidance and control of depth. Finally, the tail section contains the rudders, exhaust valves and propellers. Orders of a torpedo electric are similar to those of steam torpedoes, but the tank air is replaced by batteries and the turbines by an electric motor [1-2].

The sliding mode control has proved its effectiveness through the theoretical studies. Its principal scopes of application are robotics and the electrical engines [3-10]. The advantage of such a control is its robustness and its effectiveness versus the disturbances and the model uncertainties. Indeed, to make certain the convergence of the system to the desired state, a high level control is often requested. In addition, the discontinuous part of the control generates the chattering phenomenon which is harmful for the actuators. In fact, there are many solutions suggested to this problem. In literature, sliding mode control with limiting band has been considered by replacing the discontinuous part of the control with a saturation function [11]. Also, fuzzy control was proposed as a solution thanks to its robustness. In another hand, the high order sliding mode consists in the sliding variable system derivation [12]. This method allows the total rejection of the chattering phenomenon while maintaining the robustness of the approach. For this approach, two algorithms could be used:

✓ the twisting algorithm: the system control is increased by a nominal control $u_e$; the system error, on the phase plane, rotates around the origin until been cancelled. If we derive the sliding surface (S) n times we see that the convergence of S is even more accurate when n is higher.

✓ the super twisting algorithm: the system control is composed of two parts $u_1$ and $u_2$ with $u_1$ equivalent control and $u_2$ the discontinuous control used to reject disturbances. In this case, there is no need to derive the sliding surface. To obtain a sliding mode of order n, in this method, we have to derive the error of the system n times [13-16].

In the literature, different approaches have been proposed for the synthesis of nonlinear surfaces. In [17], the proposed area consists of two terms, a linear term that is defined by the Herwitz stability criteria and another nonlinear term used to improve transient performance. In [18], to measure the armature current of a DC motor, Zhang Li used the high order sliding mode since it is faster than the traditional methods such as vector control ... To eliminate the static error that appears while parameters measurements one use a P.I controller [19]. Thus the author have chosen to write the sliding surface in a transfer function of a proportional integral form while respecting the convergence properties of the system to this surface. The same problem of the static error was also treated by adding an integrator block just after the sliding mode control.

## 2. PROCESS MODELING

Torpedoes (Figure 1) are systems with strong non linearity and always subject to disturbances and model parameters uncertainties which makes their measurement and their control a hard task. Equation (1) represents the torpedo's motion's equation in 6 degrees of freedom. **M** is the matrix of inertia and added inertia, **C** is the matrix of Coriolis and centrifugal terms, **D** is the matrix of hydrodynamic damping terms, **G** is the vector of gravity and buoyant forces, and $\tau$ is the control input vector describing the efforts acting on the torpedo in the body-fixed frame. **B** is a nonlinear function depending of the actuators characteristics, and **u** is the control-input vector [2].

$$M\dot{v} + C(v)v + D(v)v + G(\eta) = \tau$$
$$\tau = B(u)$$
(1)

For the modelling of this system, two references are defined (Figure 2): one fix reference related to the vehicle which defined in an origin point: $R_0$ ($X_0$, $Y_0$, $Z_0$), the second one related to the Earth R(x, y, z). The torpedo present a strong nonlinear aspect that appears when we describe the system in 3 dimensions (3D), so the state function will present a new term of disturbances as shown in (2).

$$\dot{X} = AX + Bu + \varphi(X, u)$$
(2)

with $\|\varphi(X,u)\| \le MX$, M>0.

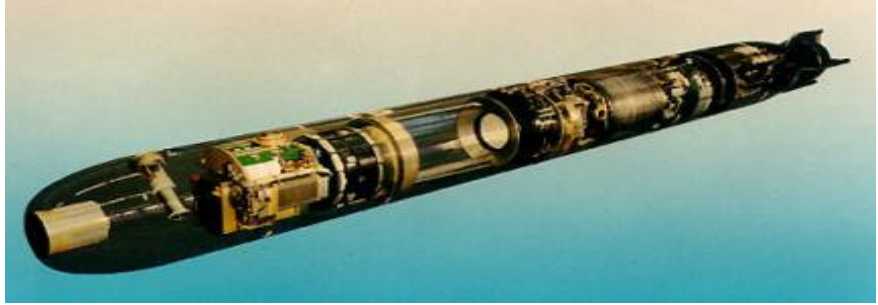

Figure 1 Schematic of a Torpedo

As we consider only the linear movement in immersion phase, we need only four degree of freedom for that we describe the system only in 2 dimensions (2 D). All development done, the resulting state space describing the system is given by (3) and (4).

$$\dot{X} = AX + Bu \qquad (3)$$

$$\dot{X} = \begin{bmatrix} \dot{\omega} \\ \dot{q} \\ \dot{\theta} \\ \dot{z} \end{bmatrix}, \quad A = \begin{bmatrix} a_{11} & a_{12} & 0 & 0 \\ a_{21} & a_{22} & a_{23} & 0 \\ 0 & 1 & 0 & 0 \\ 1 & 0 & a_{43} & 0 \end{bmatrix} \quad and \quad B = \begin{bmatrix} b_{11} \\ b_{21} \\ 0 \\ 0 \end{bmatrix} \qquad (4)$$

Where:
**ω** is linear velocity, **q** the angular velocity, **θ** the angle of inclination and **z** the depth. The system control is provided by: **u** which presents the immersion deflection.

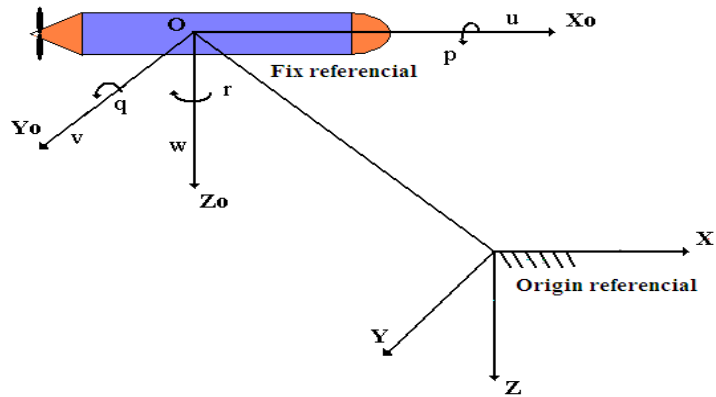

Figure 2 Inertial frame & body-fixed frame

In this way, the system could be represented by two parts [2]: $H_1(p)$ the transfer function of inclination (5) and $H_2(p)$ the transfer function of immersion (6).

$$H_1(p) = \frac{7660}{p(p+40)} \qquad (5)$$

$$H_2(p) = \frac{6514(p+6.85)}{p(p+1.91)(p+12.5)(p+40)} \tag{6}$$

## 2. THE TORPEDO CONTROLLER DESIGN

### 3.1 The sliding mode control

The appearance of the sliding mode approach occurred in the Soviet Union in the Sixties with the discovery of the discontinuous control and its effect on the system dynamics. This approach is classified in the monitoring with Variable System Structure (VSS). The sliding mode is strongly requested seen its facility of establishment, its robustness against the disturbances and models uncertainties. The principle of the sliding mode control is to force the system to converge towards a selected surface and then to remain there and to slide on in spite of uncertainties and disturbances [20-24]. The surface is defined by a set of relations between the system variables state. The synthesis of a control law by sliding mode includes two phases:
- the sliding surface is defined according to the control objectives and to the wished performances in closed loop,
- the synthesis of the discontinuous control is carried out in order to force the system state trajectories to reach the sliding surface, and then, to evolve in spite of uncertainties, of parametric variations,… the sliding mode exists when commutations took place in a continuous way between two extreme values $u_{max}$ and $u_{min}$. To ensure a good commutation, we choose a relay type control, we gets the desired result when commutations are sufficiently high. The sliding mode control has largely proved its effectiveness through the reported theoretical studies. Its principal scopes of application are robotics and the electrical motors.

For any control device which has imperfections such as delay, hystereses, which impose a frequency of finished commutation, the state trajectory oscillate then in a vicinity of the sliding surface. A phenomenon called chattering appears.

In general idea, the main purpose of the sliding mode control consists in bringing back the state trajectory towards the sliding surface and to make it move above this surface until reaching the equilibrium point. The sliding mode exists when commutations between two controls $u_{max}$ and $u_{min}$ remains until reaching the desired state. In another hand, the sliding mode exists when: $s\dot{s} < 0$. This condition is based on Lyapunov quadratic function. In fact, control algorithms based upon Lyapunov method have proven effectiveness for controlling linear and nonlinear systems subject to disturbances. In this way, the existence condition of sliding mode control can be satisfied by the candidate Lyapunov function: $V = \frac{1}{2}s^2$.

There are three different sliding mode structures: first, commutation takes place on the control unit (Figure 3), the second structure uses commutation on the state feedback (Figure 4) and finally, it is a structure by commutation on the control unit with addition of the equivalent control (Figure 5).

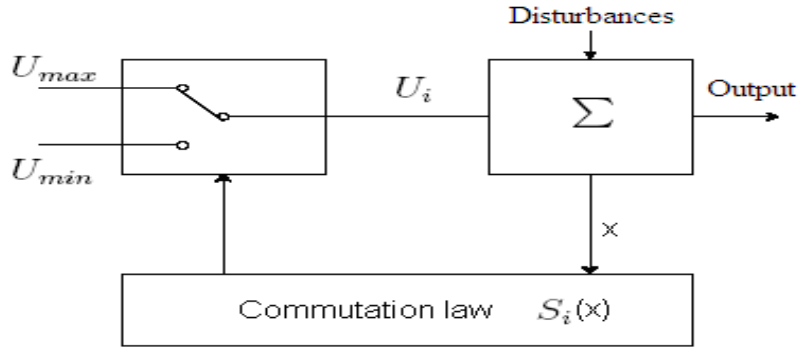

Figure 3 Control unit commutation structure

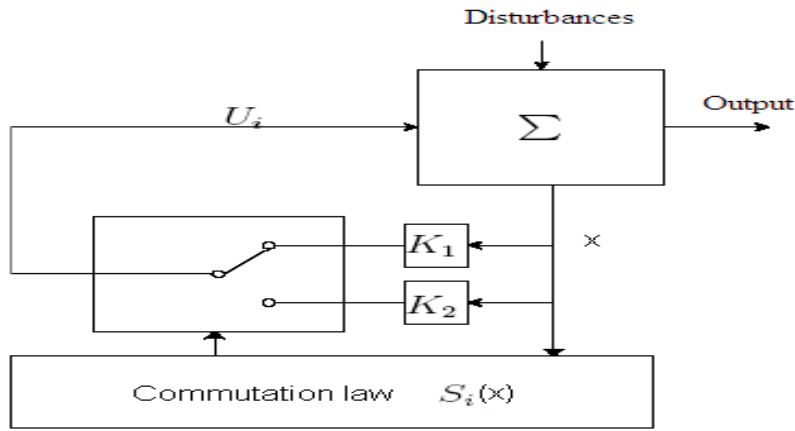

Figure 4 Control unit with commutation on the state feedback

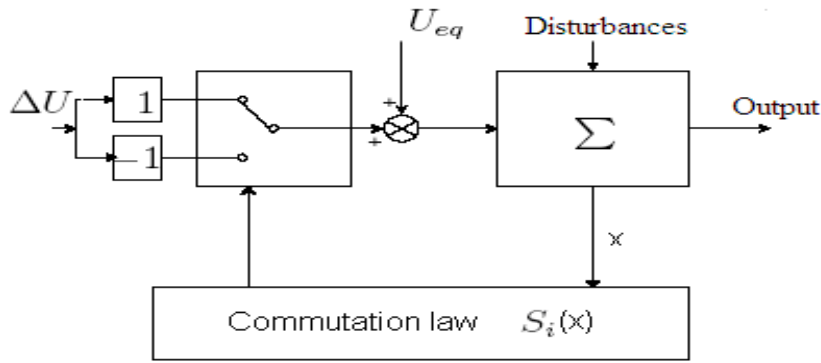

Figure 5 Control unit with addition of the equivalent control

In this study we chose to use the first structure because it's the most solicited (Figure.3). To ensure the existence of the sliding mode, we must produce a high level of discontinuous control. For that we will use a relay which commutates between two extreme values of control.

Second, we have to define a first order sliding surface. In this study we will describe it as follow (7):

$$s = k_1 e + k_2 \dot{e} \tag{7}$$

In the convergence phase to the sliding surface, we have to verify that:

$$\dot{V} = \frac{1}{2}\frac{\partial}{\partial t}(s^2) \leq -\eta |s| \tag{8}$$

with $\eta > 0$

In this part of controlling, the control law of the sliding mode could be given by (9).

$$u = k\ sign(s) \tag{9}$$

with :

$$sign(s) = \begin{cases} 1 &, s > 0 \\ -1 &, s < 0 \end{cases} \tag{10}$$

*sign(.)*: is the sign function
*k* : a positive constant that represent the discontinuous control gain

*Chattering phenomenon*

The sliding mode control has been always considered as a very efficient approach. However, considered that it requires a high level frequency of commutation between two different control values, it may be difficult to put it in practice.

In fact, for any control device which presents non linearity such as delay or hysteresis, limited frequency commutation is often imposed, other ways, the state oscillation will be preserved even in vicinity of the sliding surface. This behaviour is known by chattering phenomenon.

This highly undesirable behaviour may excite the high frequency unmodeled dynamics which could result in unforeseen instability, and can cause damage to actuators or to the plant itself. In this case the high order sliding mode can be a solution.

### 3.2 High order sliding mode control synthesis

As described before, the high order sliding mode control can be represented by two different methods: the twisting and the super twisting algorithms. In [23], a comparison between the two algorithms was achieved and notes that the super twisting algorithm is more reliable than the twisting algorithm, since it does not ensure the same robustness to perturbations. Indeed, in his article [23], the author used the second order sliding mode to improve the performances of a turbine torque. View that the conventional control approaches, that of double-fed asynchronous generator and predefined equations proved incapable of providing a convergence of the torque to the desired value. Then the choice of the high order sliding mode approach was based on its robustness against the disturbances. On the other hand, the use of linear surfaces in the control laws synthesis by sliding mode is considered satisfactory, by authors, in terms of stability. However, the dynamics imposed by this choice is relatively slow. To overcome this problem, we may use nonlinear sliding surfaces. In the same direction, to work on the speed and position regulation or power of asynchronous machines, we often use to limit the stator current (torque) that can damage the system. In this case, the authors suggested the use of the high order sliding mode approach considering a nonlinear switching law that consists of two different sliding surfaces $S_1^+$ and $S_1^-$ using two switched position. Thus, we get two limits bands, a lower band and a higher one that reduces the chattering phenomenon.

In the literature, different approaches have been proposed for the synthesis of nonlinear surfaces. In [21], the proposed area consists of two terms, a linear term that is defined by the Herwitz stability criteria and another nonlinear term used to improve transient performance. In [10], to measure the armature current of a DC motor, Zhang Li used the high order sliding mode since it is faster than traditional methods such as vector control ... To eliminate the static error that appears when measuring parameters we use a P.I controller. Thus the author have chosen to write the sliding surface in a transfer function of a proportional integral form while respecting the convergence properties of the system to this surface. The same problem of the static error was treated by adding an integrator block just after the sliding mode control [23-30].

The tracking problem of a torpedo is treated by using sliding mode control with nonlinear sliding surface as shown in (7). Consider a non linear monovariable and uncertain system characterized by the system (11).

$$\dot{x} = f(x) + g(t,x)u$$
$$y = h(x) \tag{11}$$

where $x = [x_1 \ldots x_n]^T \in X \subset IR^n$ represent the state system with X an open of $IR^n$, and $u \in U \subset IR^n$ the input of system control. We suppose that the control input $u$ is limited. The system output is represented by $y=h(x) \in Y \subset IR$ with Y an open of *IR*. *f(x)*, *g(x)* and *h(x)* are differentiable known functions.

The aim of the high order sliding mode control is to force the system trajectories to reach in finite time on the sliding ensemble of order $r \geq \rho$ defined by:

$$S^r = \{x \in IR^n : s = \dot{s} = \ldots = s^{(r-1)} = 0\}, r \in IN \tag{12}$$

$\rho > 0$, $s(x,t)$ the sliding function : it is a differentiable function with its *(r - 1)* first time derivatives depending only on the state *x(t)* (that means they contain no discontinuities). In the case of second order sliding mode control, the following relation must be verified:

$$s(t,x) = \dot{s}(t,x) = 0 \tag{13}$$

The derivative of the sliding function is

$$\frac{d}{dt}s(t,x) = \frac{\partial}{\partial t}s(t,x) + \frac{\partial}{\partial x}s(t,x)\frac{\partial x}{\partial t} \tag{14}$$

Considering relation (13) the following equation can be written as:

$$\dot{s}(t,x,u) = \frac{\partial}{\partial t}s(t,x) + \frac{\partial}{\partial x}s(t,x)\dot{x}(t) \tag{15}$$

The second order derivative of *S(t,x)* is :

$$\frac{d^2}{dt}s(t,x,u) = \frac{\partial}{\partial t}\dot{s}(t,x,u) + \frac{\partial}{\partial x}\dot{s}(t,x,u)\frac{\partial x}{\partial t} + \frac{\partial}{\partial u}\dot{s}(t,x,u)\frac{\partial u}{\partial t} \tag{16}$$

This last equation can be written as follows:

$$\frac{d}{dt}\dot{s}(t,x,u) = \xi(t,x) + \psi(t,x)\dot{u}(t) \tag{17}$$

with:

$$\xi(t,x) = \frac{\partial}{\partial t}\dot{s}(t,x,u) + \frac{\partial}{\partial x}\dot{s}(t,x,u)\dot{x}(t) \tag{18}$$

$$\psi(t,x) = \frac{\partial}{\partial u} \dot{s}(t,x,u) \tag{19}$$

We consider a new system whose state variables are the sliding function $s(t,x)$ and its derivative $\dot{s}(t,x)$.

$$\begin{cases} y_1(t,x) = s(t,x) \\ y_2(t,x) = \dot{s}(t,x) \end{cases} \tag{20}$$

Eqs. (17) and (20) lead to (21)

$$\begin{cases} \dot{\omega}_1(t,x) = \omega_2(t,x) \\ \dot{\omega}_2(t,x) = \xi(t,x) + \psi(t,x)\dot{u}(t) \end{cases} \tag{21}$$

In this way a new sliding function $\sigma(t,x)$ is proposed:

$$\sigma(t,x) = \alpha_2 \omega_2(t,x) + \alpha_1 \omega_1(t,x) = \alpha_2 \dot{s}(t,x) + \alpha_1 s(t,x) \tag{22}$$

with $\alpha_i > 0$

Eqs. (7) and (22) leads to (23)

$$\sigma = \beta_1 e + \beta_2 \dot{e} + \beta_3 \ddot{e} \tag{23}$$

with $\beta_i > 0$

In this way, using the P.I.D controller, the sliding surface will be represented as written below (24).

$$s = \alpha_1 e + \alpha_2 \dot{e} + \alpha_3 \int_0^t e \, dt \tag{24}$$

then,

$$\dot{s} = \alpha_1 \dot{e} + \alpha_2 \ddot{e} + \alpha_3 e \tag{25}$$

To reduce the chattering phenomenon, we will use the saturation function which gives:

$$u = \lambda sat\left(\frac{s}{\phi}\right) \tag{26}$$

where,

$$u = \begin{cases} \lambda sign(s) & if \ |s| \geq \phi \\ \lambda \left(\dfrac{s}{\phi}\right) & if \ |s| \leq \phi \end{cases} \tag{27}$$

with $\lambda$ and $\phi > 0$, $\phi$ defines the thickness of the boundary layer.

## 4. SIMULATION RESULTS

Simulations results are illustrated in figures 6 to 17. Figure 6 and 7 shows that, with a first order sliding mode control (SMC1) using the sliding surface (7) with $k_1 = 1$ and $k_2 = 2.5$, we can reach the desired value in a short time but the control level $(u = \pm 3)$ and its switching frequency are high (Figure 8). In addition, we notice that the reaching phase present some commutations known as chattering effect. However, the second order sliding mode control (SMC2), with the sliding surface (23) coefficients $\beta_1 = 2$, $\beta_2 = 5$, $\beta_3 = 2$, can reduce considerably the chattering phenomenon (Figure 10 and 11) but the level of the control is always high $(u \approx \pm 1.8)$ and its commutation frequency is even higher (Figure 12).

As a solution to this problem, we apply the PID-SMC1 which sliding surface defined in (24) with $\alpha_1 = 1$, $\alpha_2 = 4$ and $\alpha_3 = 0.04$. To reach the sliding surface and to converge to zeros, we choose $\phi = 2$ and $\lambda = 1$. The simulation results of this approach are given in Figure 14-17. We notice that the system error converges to zero (Figure 14 and 15) and that we have reduced considerably the chattering effect relatively to the two last approaches simulated in this paper. Other ways, we notice that the control level (Figure 16) has little commutation in the beginning of the system evolution then it stabilizes in $(u = 0.4)$ after a short period of time (~30s). This excellent result is also validated in figures 9, 13 and 17 which present the pitching angle of the torpedo which seems very adequate with the PID-SMC1 approach (figure 17).

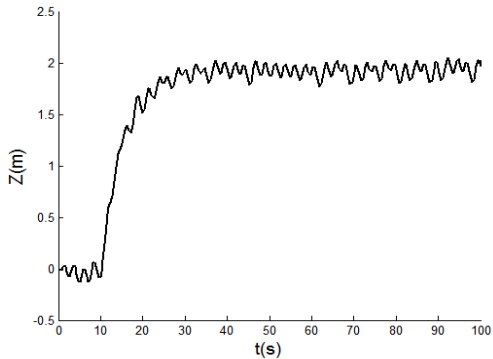
Figure 6 System immersions by SMC1

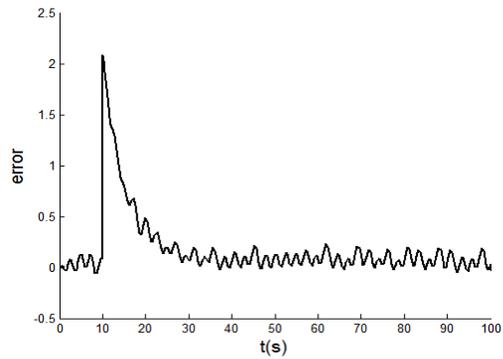
Figure 7 System error by SMC1

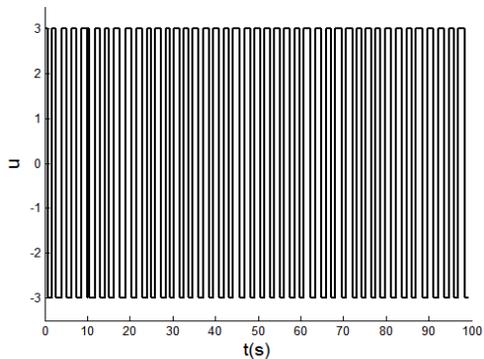
Figure 8 Control evolutions by SMC1

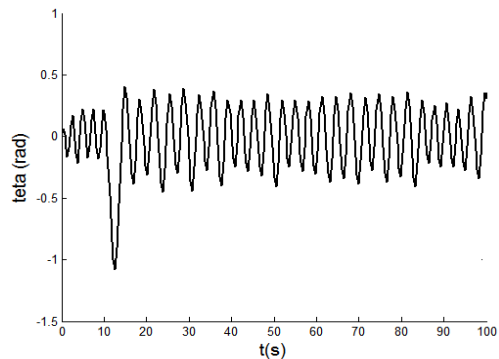
Figure 9 Immersion angle by SMC1

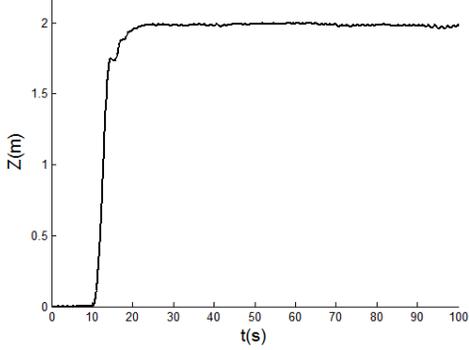

Figure 10 System immersions by SMC2

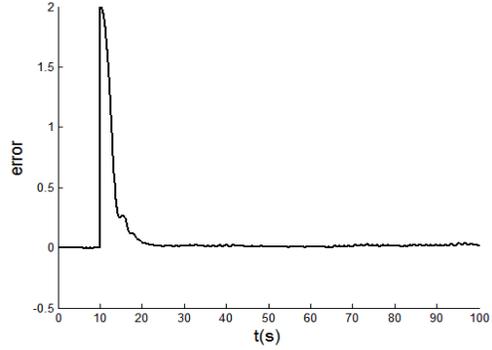

Figure 11 System error by SMC2

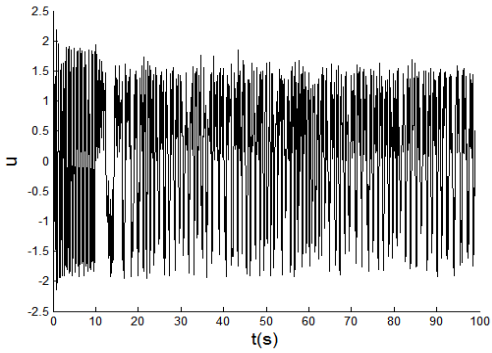

Figure 12 Control evolutions by SMC2

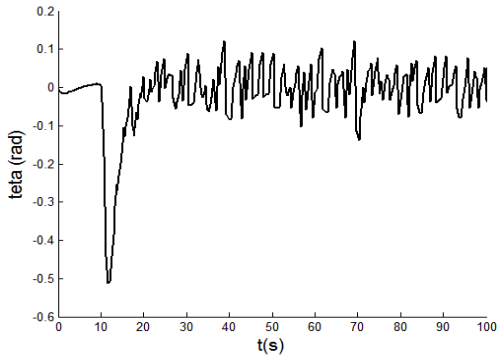

Figure 13 Immersion angle by SMC2

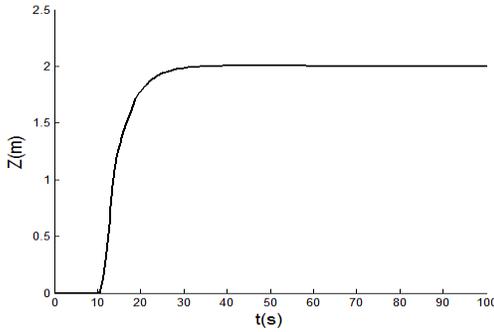

Figure 14 System immersions by PID-SMC1

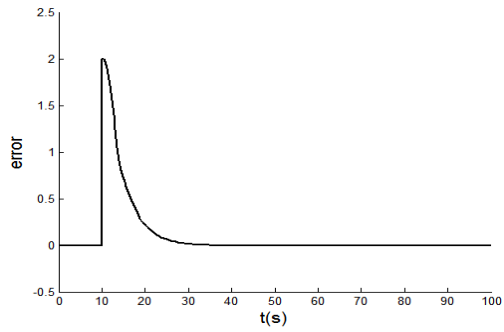

Figure 15 System error by PID-SMC1

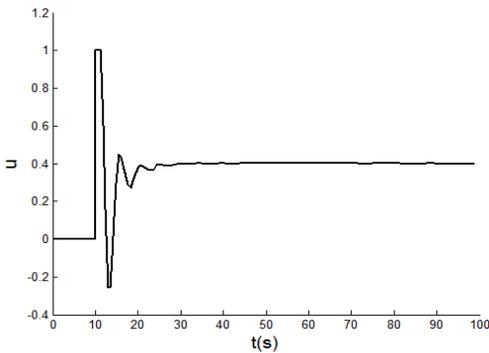

Figure 16 Control evolutions by PID-SMC1

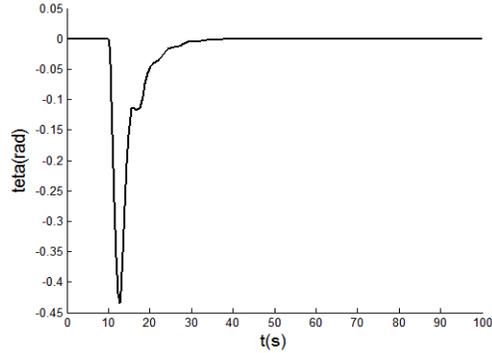

Figure 17 Immersion angle by PID-SMC1

## 5. CONCLUSION

In this work, torpedoes controller have been presented, detailed and justified by simulation results. This paper could be a ready study for those who want to start research with the sliding mode control. We approached the synthesis method of a control law by sliding mode using a nonlinear sliding surface. In the first time, we presented the class and the properties of this sliding surface adopted. Then, a sliding mode control using the sliding surface developed together with stability studies were elaborated. After that, second order sliding mode control was developed and tested by simulation on a torpedo. Finally, to reduce the static error and the number of derivative which makes the system inaccurate and which could represent some imperfections in real applications, a PID sliding surface had been developed and simulated. This last approach show very effective qualities of control and robustness especially in term of the control level reduction and the sliding mode switching control minimization.

International Journal of Information Technology, Control and Automation (IJITCA) Vol.2, No.1, January 2012

**Author**

**Ahmed Rhif** was born in Sousah, Tunisia, in August 1983. He received his Engineering diploma and Master degree, respectively, in Electrical Engineering in 2007 and in Automatic and Signal Processing in 2009 from the National School of Engineer of Tunis, Tunisia (L'Ecole Nationale d'Ingénieurs de Tunis E.N.I.T). He has worked as a Technical Responsible and as a Project Manager in both LEONI and CABLITEC (Engineering automobile companies). Then he has worked as a research assistant at the Private University of Sousah (Université Privée de Sousse U.P.S) and now at the High Institute of Applied Sciences and Technologies of Sousah (Institut Supérieur des Sciences Appliquées et de Technologie de Sousse I.S.S.A.T.so). He is currently pursuing his PhD degree in the Polytechnic School of Tunis (E.P.T). His research interest includes control and nonlinear systems.
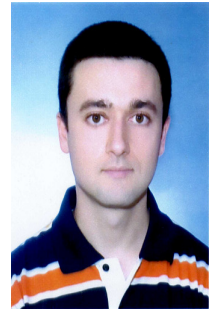
**E-mail :** ahmed.rhif@gmail.com